\documentclass[11pt]{article}
\usepackage[dvips]{graphicx}  
\begin{document}

\title{\bf 
Gravitation, the `Dark Matter'
 Effect 
and the Fine Structure Constant}

\author{{ \bf Reginald T. Cahill}\\
School of Chemistry, Physics and Earth Sciences\\
 Flinders University \\
 GPO Box 2100, Adelaide 5001, Australia \\
 Reg.Cahill@flinders.edu.au \\ \\ \\physics/0401047\\\\  
Published in {\it Apeiron} Vol. 12, No. 2, April 2005, 144-177.}
\date{}

\maketitle

\begin{center} Abstract \end{center}
Gravitational anomalies such as  the mine/borehole $g$
anomaly, the near-flatness of the spiral galaxy rotation-velocity curves, currently interpreted as the  `dark
matter' effect, the absence of that  effect in ordinary elliptical galaxies, and  the ongoing
problems in accurately determining Newton's gravitational constant $G_N$ are explained by a  generalisation of 
the Newtonian 
theory of gravity  to   a fluid-flow  formalism with one new dimensionless constant.  By analysing the borehole   
data this  new
constant is shown to be the fine structure constant $\alpha \approx 1/137$.  The spiral galaxy 
rotation curve effect and
the globular cluster central `black hole' masses for M15 and G1 are then correctly predicted.

\vspace{10mm}

\newpage

\tableofcontents

\vskip12pt
\section{Introduction\label{section:introduction}}
\vskip6pt

Gravity has played a key role in the history of  physics, with first the successes of the 
Newtonian theory and later the putative successes of
the Einsteinian theory, General Relativity.  
However there are numerous gravitational phenomena which are inexplicable
 within both the Newtonian and Einsteinian theories of gravity,
including the mine/borehole
$g$ anomaly \cite{Stacey,Holding,Greenland}, the almost flat rotation-velocity curves of spiral galaxies \cite{URC}, the
absence of that effect in ordinary elliptical galaxies \cite{Elliptical}, and an ongoing lack of convergence in measurements
of the Newtonian Gravitational constant $G_N$ over the last 60 years \cite{Gvariations}, and other anomalies not discussed here. The spiral
galaxy effect has been interpreted as being caused by an unknown form of `dark matter' \cite{Freeman}. 
  
It would at first appear highly unlikely that a new theory of gravity 
could supersede General Relativity by passing the same tests and yet explaining also
the various anomalies. However this is the situation that is now unfolding.  
The decisive  tests of General Relativity  were in 
situations where the external Schwarzschild metric was applicable,
namely external to a spherically symmetric matter distribution. A critical insight  is that the gravitational anomalies involve either a
non-spherical matter distribution, as in spiral galaxies, or are internal to a spherical matter distribution, as for the borehole anomaly.
  It turns out that the Newtonian theory can
be exactly re-written in the language of a `fluid in-flow' system.    Historically the Newtonian theory
was based on observations within the solar system, in which small test `objects', planets, are in orbit about a large
central mass - the sun. This led to Newton's famous inverse square law, where the gravitational force is inversely proportional to the square of
the distance.  In the new theory of gravity this law turns out to be only valid  under special conditions. In other cases the
gravitational force is different to that from Newtonian gravity.  The evidence is that there exists a non-Newtonian aspect to gravity
even in the non-relativistic limit. 

The new  generalised `fluid in-flow' formalism involves one new dimensionless constant, so that now gravity involves
two constants, this new constant and the familiar $G$.  The surprising discovery reported herein is that this new
constant is none other than the fine structure constant $\alpha=e^2/\hbar c= 1/137.036$.  This discovery suggests
that space has a quantum structure,   even though the flow equation is itself a classical equation, i.e., the quantum
effects are apparent at the classical level.  The occurrence of $\alpha$ does not necessarily imply that it is
Quantum Electrodynamics (QED) that is playing a role.  In QED $\alpha$ plays the role of the probability  of charged
particles to emit/absorb a photon, and it is probably this role which is being now revealed as a generic role for
$\alpha$, namely that it is a generic measure of randomness at a very fundamental  level.  If this interpretation is valid then it 
suggests that the gravitational anomalies were then really quantum gravity effects.  In gravity theories involving only $G$ it was
expected that quantum gravity effects would only show up at the scale of the Planck length, ${\it l}_P=\sqrt{\hbar G/c^3}\approx
10^{-35}$m, and time, ${\it t}_P=\sqrt{\hbar G/c^5}\approx 10^{-44}$s , but this may now turn out to have been an incorrect
conjecture. Quantum gravity effects may in fact be relatively large and easily observed, just as they are in atomic systems. Indeed as
discussed herein the Cavendish-type laboratory experiments have revealed systematic discrepancies of the order of
$\alpha/4$, and so now a new analysis of data from such experiments  is capable of  giving the value of
$\alpha$ via purely laboratory gravity experiments.

One new implication of the theory is that it successfully predicts  the masses of the  `black holes' that
 have recently been reported
at the centres of globular clusters, and this phenomenon also involves the value of $\alpha$.   So it 
turns out that both the Newtonian and
Einsteinian theories of gravity are  only valid in very special cases, and it was from these  cases that
these theories were incorrectly judged to  offer an explanation of gravitational phenomena.

Here  we derive the  `in-flow' theory of gravity, which 
   involves a  classical velocity field  and the theory  exhibits  the `dark matter'
effect, with strength set by the fine structure constant.
This flow theory is apparently the classical description of a quantum foam substructure to space , and the `flow' describes
the relative motion of this quantum foam with, as we now show, gravity arising from inhomogeneities and time variations in that
flow. These gravitational effects can be caused by an in-flow into matter, or even  produced purely by the
self-interaction of space itself, as happens for instance  for the new `black holes', which do not contain in-fallen 
matter. 

\vskip12pt
\section[Gravity and the `Dark Matter'  Effect]{\bf   Gravity and the `Dark Matter' \newline Effect} 
\vskip6pt

The apparently most successful theory of gravity is the Einstein General Relativity (GR)  which supposes a 4-dimensional
differential manifold with a metric tensor  $g_{\mu\nu}(x )$ which specifies the proper time interval according to
\begin{equation}
d\tau^2= g_{\mu\nu}(x )dx^\mu dx^\nu.
\label{eqn:A0}\end{equation}
Trajectories of test objects  are determined by extremising the proper time $\delta \tau/\delta x_\mu=0$, giving the
geodesic equation in terms of the usual affine connection, constructed from $g_{\mu\nu}(x )$,
\begin{equation}
\Gamma^\lambda_{\mu\nu}\frac{dx^\mu}{d\tau}\frac{dx^\nu}{d\tau}+\frac{d^2x^\lambda}{d\tau^2}=0.
\label{eqn:geodesic}\end{equation}
However all direct tests or observations of the GR formalism have used only the external Schwarzschild metric, for which
(\ref{eqn:A0}) takes the well-known form
\begin{equation}
d\tau^2=(1-\frac{2GM}{c^2r})dt^{ 2}-
\frac{1}{c^2}r^{ 2}(d\theta^2+\sin^2(\theta)d\phi^2)-\frac{dr^{ 2}}{c^2(1-\frac{\displaystyle
2GM}{\displaystyle c^2r})}.
\label{eqn:A1}\end{equation}
external to a spherical mass $M$. However by way of the change of variables 
$t\rightarrow t^\prime$ and
$\bf{r}\rightarrow {\bf r}^\prime= {\bf r}$ with
\begin{equation}
t^\prime=t+
\frac{2}{c}\sqrt{\frac{2GMr}{c^2}}-\frac{4GM}{c^2}\mbox{tanh}^{-1}\sqrt{\frac{2GM}{c^2r}},
\label{eqn:37}\end{equation}
(\ref{eqn:A1})  may be written in the form
\begin{equation}
d\tau^2=dt^{\prime 2}-\frac{1}{c^2}(dr^\prime+\sqrt{\frac{2GM}{r^\prime}}dt^\prime)^2-\frac{1}{c^2}r^{\prime
2}(d\theta^{\prime 2}+\sin^2(\theta^\prime)d\phi^{\prime 2}),
\label{eqn:PG}\end{equation}
with $r^\prime$ is the radial distance, and
which involves  the radial in-flow velocity field
\begin{equation}
{\bf v}({\bf r})=-\sqrt{\frac{2GM}{r}}\hat{\bf r}.  
\label{eqn:A2}\end{equation}
So in all cases the explicit tests of GR actually involved a velocity field.  Cases where the metric is not equivalent to
(\ref{eqn:A1}) or (\ref{eqn:PG}) have not been experimentally tested. 
This and other experimental evidence, see below, suggest that gravity may be in fact a consequence of a flow field, and
that the metric formalism may have been misleading. A form for the proper time for a general velocity field  ${\bf v}({\bf
r}(t),t)$, that generalises (\ref{eqn:PG}), is
\begin{equation}
d\tau^2=g_{\mu\nu}dx^\mu dx^\nu=dt^2-\frac{1}{c^2}(d{\bf r}(t)-{\bf v}({\bf r}(t),t)dt)^2.
\label{eqn:PGmetric}\end{equation}
Then the geodesic equation  (\ref{eqn:geodesic}) is explicitly computed to give the acceleration of the test object
\begin{equation}\label{eqn:A3}
 \frac{d {\bf v}_0}{dt}=
\left(\displaystyle{\frac{\partial {\bf v}}{\partial t}}+({\bf v}.{\bf \nabla}){\bf
v}\right)+({\bf \nabla}\times{\bf v})\times{\bf v}_R-\frac{{\bf
v}_R}{1-\displaystyle{\frac{{\bf v}_R^2}{c^2}}}
\frac{1}{2}\frac{d}{dt}\left(\frac{{\bf v}_R^2}{c^2}\right),
\end{equation}
where ${\bf v}_0$ is the velocity of the test object, and ${\bf v}_R({\bf r}(t),t)={\bf v}_0-{\bf v}({\bf r}(t),t)$ is the
velocity of the test object relative to the local `substratum' that actually is flowing, according to the frame  to which
positions and speeds are referenced. To be explicit the  frame defined by the Cosmic Background Radiation (CBR) could be used,
though this does not imply any special local privilege to the frame. Eqn.(\ref{eqn:A3}) is exact for metrics of the form in
(\ref{eqn:PGmetric}), which are known as Panlev\'{e}-Gullstrand metrics. Of course (\ref{eqn:A3}) is independent of the mass of the
test object, which is the equivalence principle. Eqn.(\ref{eqn:A3}) is particularly revealing. The first term is the well-known Euler
`total derivative' fluid acceleration, and involves the explicit time-dependence as well as the convective fluid acceleration
component, the 2nd term is the  Helmholtz fluid acceleration component caused by vorticity in the flow, while the last term is the
relativistic effect, which causes precession of elliptical orbits, event horizons, etc.  This form then suggests that the phenomenon 
of gravity is caused by time variations and inhomogeneities of some flow, and that the curved spacetime manifold mathematics was
essentially concealing that observation.  This of course suggests a critical reassessment even of the Newtonian gravity formalism.

The Newtonian
theory was  formulated in terms of a force field, the gravitational acceleration
${\bf g}({\bf r},t)$, and was based on Kepler's laws for the observed motion of the planets within the solar
system.  Newton had essentially suggested that   
${\bf g}({\bf r},t)$ is determined by  the matter density
$\rho({\bf r},t)$ according to
\begin{equation}\label{eqn:g1}
\nabla.{\bf g}({\bf r},t)=-4\pi G\rho({\bf r},t).
\end{equation}
However the acceleration in (\ref{eqn:A3}) implies that a velocity field  formalism is more fundamental, as clearly the
acceleration cannot be re-constructed from the velocity field.  Only the terms in (\ref{eqn:A3}) independent of the
test object velocity can be dynamically associated with the flow dynamics itself, and so the Euler fluid acceleration
should be used in (\ref{eqn:g1})  in place of ${\bf g}({\bf r},t)$, giving
\begin{equation}
\nabla.\left(\frac{\partial {\bf v} }{\partial t}+({\bf
v}.{\bf \nabla}){\bf v}\right)=-4\pi G\rho,
\label{eqn:CG1}\end{equation}
with ${\bf g}$ now a derived quantity given by the Euler fluid acceleration
\begin{equation}{\bf g}({\bf r},t)=\displaystyle{\frac{\partial {\bf v}}{\partial
t}}+({\bf v}.{\bf \nabla}){\bf v}\equiv\displaystyle{\frac{d{\bf v}}{dt}},
\label{eqn:CG2}\end{equation}
the last expression defines the total Euler  fluid derivative. 
External to a spherically symmetric mass $M$ the solution to (\ref{eqn:CG1}), is (\ref{eqn:A2}), and then from
(\ref{eqn:CG2}) we get the usual inverse  square law
\begin{equation}
{\bf g}({\bf r})=-\frac{GM}{r^2}\hat{\bf r}, \mbox{\ \ }r>R.
\label{eqn:InverseSqLaw}\end{equation} 
\index{inverse square law}
It must be emphasised that the velocity field  formalism in (\ref{eqn:CG1})-(\ref{eqn:CG2}) is {\it mathematically}
equivalent to the acceleration field formalism  (\ref{eqn:g1}); they both always give the same acceleration field.  However
there are two reasons for believing that the velocity field is {\it physically} more fundamental:  (i)
(\ref{eqn:CG1})-(\ref{eqn:CG2}) permit a generalisation that leads to an explanation of the so-called `dark matter'
effect, and to numerous other effects, discussed in later sections, whereas (\ref{eqn:g1}) does not permit that
generalisation, and (ii) the velocity field has been directly observed. The experimental evidence for  the velocity field
has been extensively reported in  \cite{Cahill2,RGC}, where the velocity field is apparently associated with galactic
gravitational effects, but most significantly a smaller component of the velocity field flowing past the earth towards the
sun  has been recently extracted from  the Miller data from 1925/26, and has a value consistent with (\ref{eqn:A2}) where
$M$ is the mass of the sun.

However there is one immediate insight into gravity that arises from (\ref{eqn:CG1}), and that is that the inverse square
law for gravity is now seen to be a consequence of the  inhomogeneity part of the Euler fluid acceleration, namely
$({\bf v}.{\bf \nabla}){\bf v}$, which for zero vorticity has the form ${\bf \nabla}({\bf v}^2)/2$.  In turn the form of
this inhomogeneity is determined by the requirement that the acceleration in (\ref{eqn:CG2}) be Galilean covariant.

One  consequence of the velocity field formalism (\ref{eqn:CG1})-(\ref{eqn:CG2}) is that it can be generalised to
include a new unique term   
\begin{equation}
\frac{\partial }{\partial t}(\nabla.{\bf v})+\nabla.(({\bf
v}.{\bf \nabla}){\bf v})+C({\bf v})=-4\pi G\rho,
\label{eqn:CG3}\end{equation}
where
\begin{equation}
C({\bf v})=\displaystyle{\frac{\alpha}{8}}((tr D)^2-tr(D^2)),
\label{eqn:Cdefn1}\end{equation} and
\begin{equation}
D_{ij}=\frac{1}{2}\left(\frac{\partial v_i}{\partial x_j}+\frac{\partial v_j}{\partial x_i}\right).
\label{eqn:Ddefn1}\end{equation}
Eqn.(\ref{eqn:CG3}) has the same  solution (\ref{eqn:A2}) external to a spherically symmetric mass,  because
$C({\bf v})=0$ for that flow, and so the presence of the  $C({\bf v})$ would not have manifested in the
special case of planets in orbit about the massive central sun. So (\ref{eqn:CG3})-(\ref{eqn:CG2})  are consistent
with Kepler's laws for planetary motion in the solar system, and  including the relativistic term in 
(\ref{eqn:A3}) we obtain as well the precession of elliptical orbits. Here
$\alpha$ is a   dimensionless constant - a new gravitational constant, in addition to the   Newtonian
gravitational constant
$G$.  From (\ref{eqn:CG2}) we can write (\ref{eqn:CG3})) as
\begin{equation}\label{eqn:g2}
\nabla.{\bf g}=-4\pi G\rho-4\pi G \rho_{DM},
\end{equation}
where
\begin{equation}
\rho_{DM}({\bf r})=\frac{\alpha}{32\pi G}( (tr D)^2-tr(D^2)),  
\label{eqn:DMdensity}\end{equation} 
which introduces an effective `matter density' onto the RHS of the Newtonian formalism in (\ref{eqn:g1}), phenomenologically
representing the flow self-interaction dynamics associated with the 
$C({\bf v})$ term. However the  dynamical  effect represented by   this new term  cannot be included, in a closed form, in
the gravitational acceleration dynamics formalism of  (\ref{eqn:g1}) because it cannot be expressed in terms of the
gravitational field ${\bf g}$.   This dynamical effect is shown here to be the `dark matter' effect.  The main theme of
this paper is the determination of the value of $\alpha$ from experimental data, and then the computation of various
observed effects that then follow.

We apply the new gravity theory to an earth based experiment to determine the value of $\alpha$. However we  know that
earth in-flow  is a small component compared to the total flow, as given by the experimental data discussed in
\cite{Cahill2, RGC}. For completeness we would then need to demonstrate that the results for this experimental situation
are unaffected by the larger `background' flow. This has been done, but requires a much more detailed analysis then given
herein. Then for a zero-vorticity stationary flow, and ignoring any background flow, (\ref{eqn:CG3}) may be written in the form of a
non-linear integral equation
\begin{equation}\label{eqn:integraleqn}
{\bf v}^2({\bf r})=2G\int d^3
s\frac{\rho( {\bf s})}{|{\bf r}-{\bf s}|}+
2G\int d^3
s\frac{\rho_{DM}( {\bf v}({\bf s}))}{|{\bf r}-{\bf s}|},
\end{equation} 
as $\nabla^2 \frac{1}{|{\bf r}-{\bf s}|}=-4\pi\delta^4({\bf r}-{\bf s})  $.
In particular when the matter density and the flow are both
spherically symmetric and stationary in time (\ref{eqn:CG3}) becomes, with  $v^\prime \equiv dv/dr$, the non-linear
differential equation
\begin{equation}
2\frac{vv^\prime}{r} +(v^\prime)^2 + vv^{\prime\prime} =-4\pi G\rho(r)-4\pi G \rho_{DM}(v(r)), 
\label{eqn:InFlowRadial}
\end{equation}
with now
\begin{equation}
\rho_{DM}(v(r))= \frac{\alpha}{8\pi G}\left(\frac{v^2}{2r^2}+ \frac{vv^\prime}{r}\right).
\label{eqn:dm1}\end{equation}
Then (\ref{eqn:integraleqn}) gives a non-linear radial integral form for  (\ref{eqn:InFlowRadial}), on
doing the angle integrations,
\begin{eqnarray}
v^2(r)&=&\frac{8\pi G}{r}\int_0^r s^2 \left[\rho(s)+\rho_{DM}(v(s))\right]ds
\nonumber \\&&+8\pi G\int_r^\infty s
\left[\rho(s)+\rho_{DM}(v(s))\right]ds,
\label{eqn:integralEqn}\end{eqnarray}
It needs to be emphasised that with $\alpha=0$ (\ref{eqn:InFlowRadial}) is completely equivalent to 
Newtonian gravity. 

First  consider solutions to (\ref{eqn:dm1})  and
(\ref{eqn:integralEqn}) in the perturbative regime. Iterating once we find,
\begin{equation}
\rho_{DM}(r)=\frac{\alpha}{2r^2}\int_r^\infty s\rho(s)ds+O(\alpha^2),
\label{perturbative}\end{equation}
so that in spherical systems the `dark matter' effect is concentrated near the centre, and we find  the total `dark
matter'
\begin{eqnarray}
M_{DM}&\equiv& 4\pi\int_0^\infty r^2\rho_{DM}(r)dr=\frac{4\pi\alpha}{2}\int_0^\infty
r^2\rho(r)dr+O(\alpha^2)\nonumber \\
&=&\frac{\alpha}{2}M+O(\alpha^2),
\label{eqn:TotalDM}\end{eqnarray}
where $M$ is the total amount of (actual) matter. Hence to $O(\alpha)$   $M_{DM}/M=\alpha/2$ independently of the matter
density profile. This turns out be be directly applicable to the case of globular clusters, as shown later, and also
implies that the theory of stellar structures needs to be reconsidered, as this central `dark matter' effect changes
the central $g(r)$ considerably. This may have some bearing on the solar neutrino problem.

\vskip12pt
\section{\bf  Borehole $g$ Anomaly}
\vskip6pt 

\begin{figure}[t]
\hspace{35mm}\includegraphics[scale=0.9]{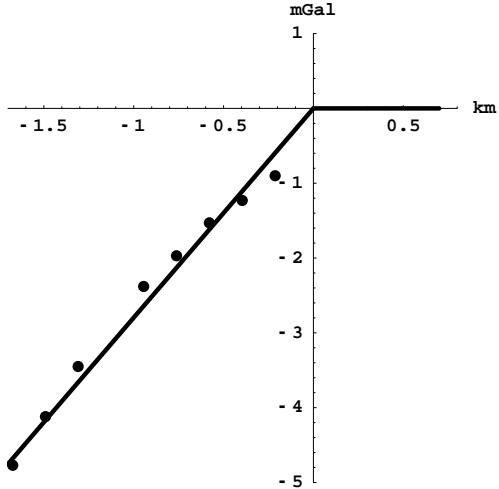}
\caption{{ The data shows the gravity residuals for the Greenland Ice Cap \cite{Greenland} Airy measurements of the
$g(r)$  profile,  defined as
$\Delta g(r) = g_{Newton}-g_{observed}$, and measured in mGal (1mGal $ =10^{-3}$ cm/s$^2$), plotted against depth in km.  Using
(\ref{eqn:deltag2}) we obtain $\alpha^{-1}=139 \pm  5 $ from fitting the slope of the data, as shown.}
\label{fig:Greenland}}\end{figure}

When the matter density $\rho(r)=0$ for $r\geq R$, as for the earth, then
 we also obtain, to $O(\alpha)$, from
 (\ref{eqn:dm1}) and (\ref{eqn:integralEqn}), and then (\ref{eqn:CG2}),
\begin{equation}
g(r)=\left\{ \begin{tabular}{ l} 
$\displaystyle{-\frac{(1+\displaystyle{\frac{\alpha}{2}}) GM}{r^2}, \mbox{\ \ } r > R,}$  \\  
$\displaystyle{-\frac{4\pi G}{r^2}\int_0^rs^2\rho(s) ds
-\frac{2\pi\alpha G}{r^2}\int_0^r\left(\int_s^R s^\prime\rho(s^\prime) ds^\prime\right) ds}$,\\$\displaystyle{
\mbox{\ \ \ \ \ \ \ \ \ \ \ \ \ \ \ \ \ \ \ \ \ \ \ \ \ \ \ \  } r
< R}$,\\ 
\end{tabular}\right.   
\label{eqn:ISL2}\end{equation}
which gives   Newton's `inverse square law' for $r > R$, but in which we see that the effective Newtonian gravitational constant
 is $G_N=(1+\frac{\alpha}{2})G$, which is  different to the fundamental gravitational constant
$G$ in (\ref{eqn:CG1}). The result in (\ref{eqn:ISL2}), which is different from that of the Newtonian theory 
($\alpha=0$) has actually been observed in mine/borehole measurements \cite{Stacey,Holding,Greenland} of
$g(r)$, though of course there had been no explanation for the effect, and indeed the
reality of the effect was eventually doubted.    
The gravity residual \cite{Stacey,Holding,Greenland}  is defined as
\begin{eqnarray}
\Delta g(r)&\equiv & g(r)_{Newton}-g(r)_{observed}\\ 
&=&g(r)_{Newton}-g(r). 
\label{eqn:deltag1}\end{eqnarray}
The `Newtonian theory' assumed in the determination of the gravity residuals  is, in the present context,
\begin{equation}
g(r)_{Newton}=\left\{ \begin{tabular}{ l} 
$\displaystyle{-\frac{G_N M}{r^2},\mbox{\ \ } r > R,}$  \\  
$\displaystyle{-\frac{4\pi G_N}{r^2}\int_0^rs^2\rho(s) ds}$,$\displaystyle{\mbox{\ \ } r < R,}$\\ 
\end{tabular}\right.   
\label{eqn:earthg2}\end{equation}
with $G_N=(1+\frac{\alpha}{2})G$. Then $\Delta g(r)$ is found to be, to 1st order in $\alpha$ and in $R-r$,  i.e.
near the surface, 
\begin{equation}
\Delta g(r)=\left\{ \begin{tabular}{ l} 
$\displaystyle{\mbox{\ \ }0, \mbox{\ \ } r> R,}$  \\   
$\displaystyle{-2\pi\alpha G_N\rho(R)(R-r),\mbox{\ \ } r < R,} $\\ 
\end{tabular}\right.    
\label{eqn:deltag2}\end{equation}
which is the form actually observed  \cite{Stacey,Holding,Greenland}. 
So outside of the spherical earth the Newtonian theory and the in-flow theory are
indistinguishable, as indicated by the horizontal line, for $r>R$, in Fig.\ref{fig:Greenland}.  However inside the earth the
two theories give a different dependence on $r$, due to the `dark matter' effect within the earth. Even though the `dark matter'
effect is concentrated near the centre in this case, there is still a small effect just beneath the surface.

  Gravity  residuals  from a borehole
into the Greenland Ice Cap  were determined  down to a depth of 1.5km \cite{Greenland}. The ice had a measured density of
$\rho=930$ kg/m$^3$, and from (\ref{eqn:deltag2}), using $G_N=6.6742\times10^{-11}$ m$^3$s$^{-2}$kg$^{-1}$, we obtain from a
linear fit to the slope of the data points in Fig.\ref{fig:Greenland} that
$\alpha^{-1}=139\pm 5$, which equals the value of the fine structure constant  $\alpha^{-1}=137.036$ to within the errors, and
for this reason we identify the  constant $\alpha$ in (\ref{eqn:Cdefn1}) as being the fine structure constant.

To confirm that this is not a coincidence we now  predict the spiral galaxy `dark matter' effect and the globular cluster
`black hole' masses using this value for $\alpha$, and also indicate the likely origin of the unexplained systematic discrepancies
apparent in the ongoing attempts to measure $G$ with increased accuracy.

\vskip12pt
\section{\bf  Spiral Galaxies}
\vskip6pt

Consider the non-perturbative solution of (\ref{eqn:CG3}), say  for a galaxy with a non-spherical matter
distribution. Then numerical techniques are necessary, but beyond a sufficiently large distance  the in-flow will
have spherical symmetry, and in that region we may use (\ref{eqn:InFlowRadial})  and (\ref{eqn:dm1}) with
$\rho(r)=0$.  Remarkably then  the pair (\ref{eqn:InFlowRadial})  and (\ref{eqn:dm1}) has an exact
non-perturbative two-parameter analytic solution,
\begin{equation}
v(r) = K\left(\frac{1}{r}+\frac{1}{R_S}\left(\frac{R_S}{r}  \right)^{\displaystyle{\frac{\alpha}{2}}}  \right)^{1/2},
\label{eqn:vexact}\end{equation}
where $K$ and $R_S$ are arbitrary constants in the $\rho=0$ region, but whose values are determined by matching to
the solution in the matter region. Here $R_S$ characterises the length scale of the non-perturbative part of this
expression,  and $K$ depends on $\alpha$ and $G$ and details of the matter distribution.  The galactic circular orbital velocities of
stars etc may be used to observe this in-flow process in a spiral galaxy and  from (\ref{eqn:CG2}) and
(\ref{eqn:vexact}) we obtain a replacement for  the Newtonian  `inverse square law' ,
\begin{equation}
g(r)=\frac{K^2}{2} \left( \frac{1}{r^2}+\frac{\alpha}{2rR_S}\left(\frac{R_S}{r}\right)
^{\displaystyle{\frac{\alpha}{2}}} 
\right),
\label{eqn:gNewl}\end{equation}
in the asymptotic limit.     From  (\ref{eqn:gNewl}) the centripetal
acceleration  relation for circular orbits 
$v_O(r)=\sqrt{rg(r)}$  gives  a `universal rotation-speed curve'
\begin{equation}
v_O(r)=\frac{K}{2} \left( \frac{1}{r}+\frac{\alpha}{2R_S}\left(\frac{R_S}{r}\right)
^{\displaystyle{\frac{\alpha}{2}}} 
\right)^{1/2}.
\label{eqn:vorbital}\end{equation}
\begin{figure}[t]
\hspace{15mm}\includegraphics[scale=1.4]{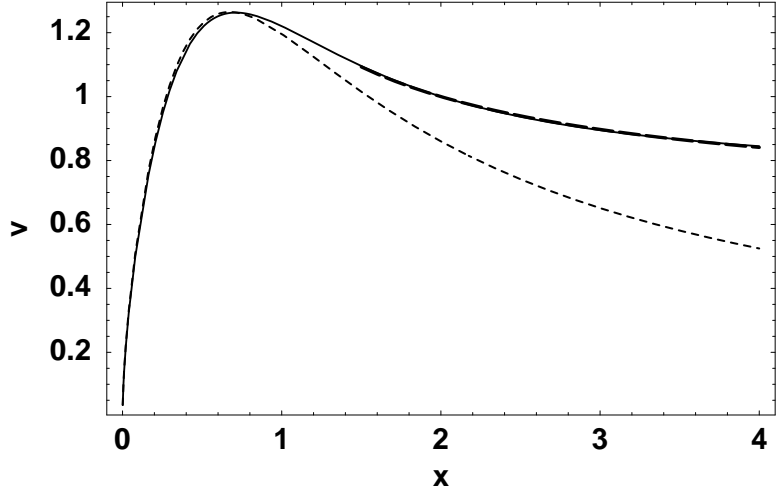}
\caption{{ Spiral galaxy rotation speed curve plots, with  $x=r/R_{opt}$.  Solid line is the Universal Rotation Curve
(URC) for  luminosity
$L/L_*=3$, using the URC in  (\ref{eqn:URC}), Ref.\cite{URC}. Short-dashes line is URC with only the matter
exponential-disk contribution, and re-fitted to the full URC at low x. Long-dashes  line, essentially overlaying the upper solid line
for $x>1.5$, is the form  in (\ref{eqn:vorbital}), for $\alpha=1/137$ and
$R_S=0.01R_{opt}$.}  
\label{fig:URCplots}}\end{figure}

\begin{figure}
\hspace{28mm}\includegraphics[scale=1.1]{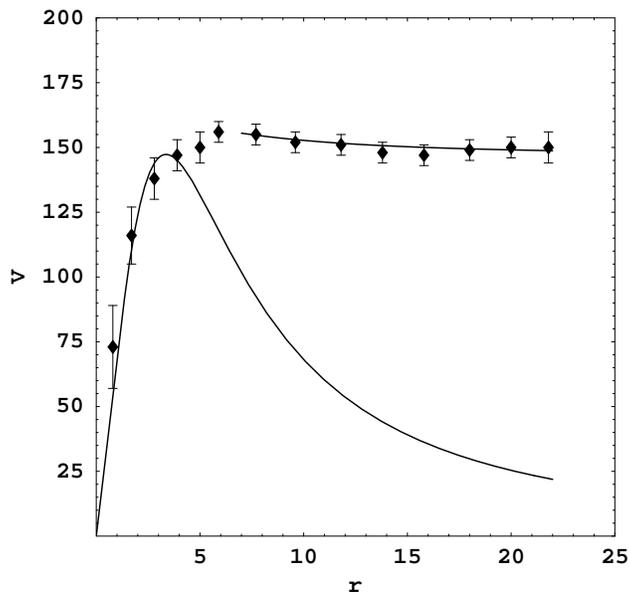}
\caption{ {Data shows the non-Keplerian rotation-speed curve $v_O$ for the spiral galaxy NGC3198 in km/s plotted against radius
in kpc/h. Lower curve is the rotation curve from the Newtonian theory or from General Relativity for an exponential disk,
which decreases asymptotically like $1/\sqrt{r}$. The upper curve shows the asymptotic form from (\ref{eqn:vorbital}), with the decrease
determined by the small value of $\alpha$.  This asymptotic form is caused by the primordial black holes at the centres of
spiral galaxies, and which play a critical role in their formation. The spiral structure is caused by the rapid in-fall
towards these primordial black holes.}
\label{fig:NGC3198}}\end{figure}

\noindent Because of the $\alpha$ dependent part this rotation-velocity curve  falls off extremely slowly with $r$, as
is indeed observed for spiral galaxies. Of course it was the inability of the  Newtonian  and Einsteinian gravity theories to explain these
observations that led to the  notion of `dark matter'. It is possible to illustrate the form in (\ref{eqn:vorbital}) by
comparing it with rotation curves of spiral galaxies.  Persic,  Salucci and   Stel
\cite{URC} analysed 
 some 1100 optical and radio rotation curves, and demonstrated that they are describable by the  empirical 
universal rotation curve (URC)
\begin{eqnarray}
v_O(x)
&=&v(R_{opt})\left[\left(0.72+0.44\mbox{Log}\frac{L}{L_*}\right)\frac{1.97x^{1.22}}{(x^2+0.78^2)^{1.43}}\right.\nonumber \\
&&\left.\mbox{\ \ \ \ \ \ \ \ \ \ \  }+1.6\mbox{e}^{-0.4(L/L_*)}
\frac{x^2}{x^2+1.5^2(\frac{L}{L_*})^{0.4}}\right]^{1/2}
\label{eqn:URC}\end{eqnarray}
where $x=r/R_{opt}$, and where $R_{opt}$ is the optical radius, or $85\%$ matter limit.  The first term is the
Newtonian contribution from an exponential matter disk, and  the 2nd term is the `dark matter' contribution. 
This two-term form also arises  from the in-flow theory, as seen in (\ref{eqn:integraleqn}).
   The form in (\ref{eqn:vorbital}) with
$\alpha=1/137$  fits, for example, the high luminosity URC, for a suitable value of $R_S$, which depends on the luminosity,
as shown by one example in Fig.\ref{fig:URCplots}.  For low luminosity data the observations do not appear to extend far enough
to reveal the asymptotic form of the rotation curve, predicted by (\ref{eqn:vorbital}).  The non-Keplerian rotation curve
effect from the new theory of gravity is shown for the spiral galaxy NGC3198 in Fig.\ref{fig:NGC3198}. 

But the general form in (\ref{eqn:vexact})
leads to a key question. Why is it that $R_S$ is essentially very large for the earth, as shown by the borehole data, and  also for
elliptical galaxies as shown
 by the recent discovery \cite{Elliptical} that planetary nebulae in ordinary elliptical 
galaxies,  serving as observable `test objects',   have Keplerian or Newtonian
rotation-speed curves, whereas spiral galaxies have small values of $R_S$ compared to their $R_{opt}$ values, and that furthermore
their
$R_S$ values are related to their luminosity.    The answer to this question is that the in-flow equation actually has a
one-parameter class of matter-free non-perturbative exact  solutions of the form
\begin{equation}
v(r) = \frac{\beta}{r^{\alpha/4}},
\label{eqn:attractor}\end{equation} 
where the $1/r$ term in (\ref{eqn:vexact}) is inadmissible because it does not satisfy the matter-free in-flow equation at $r=0$.
These solutions correspond to a novel feature of the new theory of gravity, namely the occurrence of these gravitational attractors.
These attractors presumably were produced during the big-bang, and since they can coalesce to form larger  attractors, it is most
likely that it is such an attractor that leads to the formation of spiral galaxies.    Attractors appear to form a cellular network, with
the attractor form in (\ref{eqn:attractor}) only valid for a single attractor.  Attractors with large $\beta$ values, and so large
regions of influence, will attract greater quantities of the original post-big-bang gas. As well because these have large in-flow
velocities the matter will end up with high angular momentum, resulting in a spiral galaxy.  Then the magnitude of $\beta$ is related
to the total amount of matter in the galaxy, which manifests eventually as its luminosity. Smaller attractors will form galaxies with
lower in-flow speeds and so  are less likely to have large amounts of angular momentum.  These new `gravitational
attractors' are the `black holes'  of the new theory of gravity, and their properties are determined by $\alpha$, and not
by $G$.

\vskip12pt
\section{\bf  Black Holes}
\vskip6pt 

At the center of matter distributions
the new theory of gravity also has  attractor phenomena, namely the occurrence of 
`in-flow singularities'  which, in this case, are induced by the matter, as seen in the borehole analysis.  Such in-flow
 singularities, and the `dark matter' effect in general, are mandated by the in-flow and are not contingent phenomena. These
attractor  in-flows singularities  have an event horizon, where the in-flow speed reaches the speed of light. Hence they are a new
form of `black hole'.  This phenomenon is different to that  in general relativity where  black holes arise from the past in-fall of
matter. 

Recently it has been reported  that globular clusters 
\cite{GlobularM15,GlobularG1}  have central `black holes', which now appears to be merely an interpretation of the
central `dark matter' gravitational attractor effect.   Again here the spatial structure of these `black hole' in-flow effects is
determined by
$\alpha$ - they are presumably intrinsically quantum-space processes, and the effective `mass' of this central attractor is 
computable within the new theory.    Numerical solutions of (\ref{eqn:InFlowRadial}) for typical cluster
density profiles reveal that the central  `dark matter' mass is accurately given by the perturbative result in (\ref{eqn:TotalDM}),
$M_{DM}/M=\alpha/2=0.00365$.   Then the $M_{DM}/M$ mass ratio is
independent of the density profile, as noted above.  The clusters M15 and G1 then give an excellent opportunity to test
again the new theory.  For M15 the mass of the central `black hole' was found to be \cite{GlobularM15}
$M_{DM}=1.7^{+2.7}_{-1.7}\times10^3$M$_\odot$, and the total mass of M15 was determined  \cite{M15Mass} to be
$4.9\times10^5$M$_\odot$.  Then  these results together give    $M_{DM}/M=0.0035^{+0.011 }_{-0.0035}$ which is in
excellent agreement with the above prediction. For G1 we have \cite{GlobularG1}  $M_{DM}=2.0^{+ 1.4 }_{-0.8 }\times
10^4$M$_\odot$, and $M=(7-17)\times10^6$M$_\odot$. These values give $M_{MD}/M= 0.0006-0.0049$, which is
also consistent with the above $\alpha/2$ prediction.  
There is a singularity at $r=0$ where the in-flow speed becomes unbounded, and an event horizon where $v=c$, the speed
of light. The radius of this event horizon depends on $\alpha$.   This implies that the globular cluster central
`attractor' is a manifestation of the non-Newtonian in-flow, that is, an in-flow different to the form in
(\ref{eqn:A2}).  Hence the globular cluster observations  again indicate  the role of the fine structure constant in
gravity.   

\begin{figure}
\hspace{20mm}\includegraphics[scale=1.1]{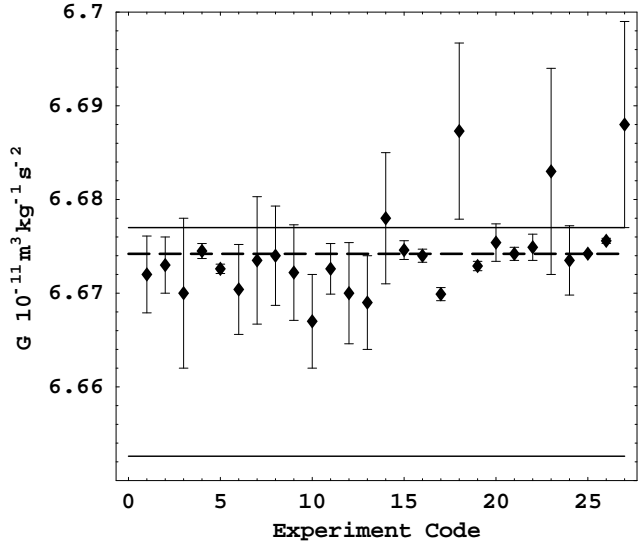}
\caption{{Results of precision measurements of $G_N$ published in the 
last sixty years in which the Newtonian theory was used to analyse the data.  
These results show  the presence of
a  systematic effect not in the Newtonian theory. 
{\bf 1:}  Gaithersburg 1942,
{\bf 2:}  Magny-les-Hameaux 1971, 
{\bf 3:}  Budapest 1974, 
{\bf 4:}  Moscow 1979,
{\bf 5:}  Gaithersburg 1982, 
{\bf 6-9:}  Fribourg   Oct 84, Nov 84, Dec 84, Feb 85,
{\bf 10:}    Braunschweig 1987,
{\bf 11:}    Dye 3 Greenland   1995,
{\bf 12:}    Gigerwald Lake 1994, 
{\bf 13-14:} Gigerwald Lake 1995  112m, 88m,
{\bf 15:}     Lower Hutt 1995    MSL, 
{\bf 16:}  Los Alamos 1997, 
{\bf 17:}    Wuhan 1998,
{\bf 18:}    Boulder JILA 1998, 
{\bf 19:}    Moscow 1998, 
{\bf 20:}    Zurich 1998, 
{\bf 21:}     Lower Hutt MSL 1999, 
{\bf 22:}    Zurich 1999, 
{\bf 23:}    Sevres 1999, 
{\bf 24:}   Wuppertal 1999,
{\bf 25:}     Seattle 2000,
{\bf 26:}     Sevres 2001, 
{\bf 27:}     Lake Brasimone 2001. 
 The upper horizontal line shows the value from the 1991 ocean measurements \cite{Ocean}, while the dashed line shows the
current  CODATA $G_N$  value based on a statistical analysis of the indicated measurements. The
lower line shows the value of $G$  after removing the `dark matter' effect within the
earth on the Ref.\cite{Ocean} $G_N$ value.}   
\label{fig:GData}}\end{figure}

\vskip12pt
\section{\bf  Measuring $G$}
\vskip6pt 

Finally it is now possible to explain the cause of the longstanding variations \cite{Gvariations} in the
measurements of the value of
$G_N$,  shown in Fig.\ref{fig:GData}. Note that the relative spread $\Delta G_N/G_N \approx O(\alpha/4)$, as we would now
expect. Essentially the different Cavendish-type experiments used different matter geometries, and as we have seen, the
 geometry of the masses  has a `non-Newtonian'  effect on the in-flow, and so on the measured force between the masses.  In these
experiments the asymptotic form in (\ref{eqn:vexact}) is not relevant as the test masses are always close, and the data indicates
non-Newtonian effects of relative size $\alpha/4$.  These effects are caused by  both a `polarisation' of the central `dark
matter' effect, caused by the presence of the other test mass, and by a `dark matter' region forming essentially between
the two masses. 

 Only for the
borehole-type experiments do we have a complete analytic analysis, and an ocean Airy measurement  of
$g$ is in this class, and 
\cite{Ocean} gives  $G_N=(6.677\pm 0.013)\times 10^{-11} $ m$^3$s$^{-2}$kg$^{-1}$, shown by the upper horizontal line in
Fig.\ref{fig:GData}.
 From that value  
 we may extract the value of the `fundamental gravitational constant' $G$ by removing the `dark matter' effect:
$G=(1-\frac{\alpha}{2})G_N+O(\alpha^2)= (6.6526 \pm 0.013)\times 10^{-11} $ m$^3$s$^{-2}$kg$^{-1}$, compared to the current CODATA
value of  
$G_N=(6.6742 \pm 0.001)\times 10^{-11}$ m$^3$s$^{-2}$kg$^{-1}$, which is contaminated with   `dark matter' effects. 
Then in the various experiments, without explicitly
computing the `dark matter' effect, one will find an `effective' value of $G_N>G$ that depends on the geometry of the masses.
A re-analysis of the data in  Fig.\ref{fig:GData} using the in-flow theory is predicted to resolve these apparent
discrepancies.  The discrepancies in measuring $G$ are then presumably quantum gravity effects and,  if so, then quantum gravity may be
easily studied  in laboratory Cavendish experiments.

\vskip12pt
\section{\bf  What Flows?}
\vskip6pt 

The evidence here is that the velocity field explanation for gravity is more encompassing of gravitational phenomena then either the
`acceleration field' theory of Newton, in the non-relativistic regime, or the `curved spacetime formalism' of Einstein. Indeed in all
cases where these two theories were successful they could be exactly recast into the velocity field formalism. But the velocity
formalism permits a unique and natural generalisation, not possible in either of these theories, and which then immediately explains
numerous so-called gravitational `anomalies', as shown herein for several examples. 

 Given that, the fundamental
question is then: what is {\it flowing}? In \cite{Cahill2,RGC,PSS} it is suggested that space has a quantum substratum, that space
is a quantum system undergoing ongoing classicalisation. As well   this quantum-foam system was argued to
arise from an {\it information theoretic} model of reality.  But what experimental  evidence is there that what flows is not some
 material moving through some space, but some very exotic and new phenomenon?  That evidence appeared  when analysing the experiments
of Michelson and Morley (1887), Miller (1925/26), and DeWitte (1991), as discussed in detail in \cite{Cahill2,RGC}. The first two
experiments were gas-mode Michelson interferometer experiments which only in 2002 were finally understood \cite{Cahill2}. Then using
this first {\it post}-relativistic effects analysis it was shown that the non-null rotation-induced fringe shifts could be understood
as arising from the combination of three effects: (i) the usual geometric path difference effect from motion through a substratum, that
Michelson had used in the design
 of his interferometer, (ii) the physical Fitzgerald-Lorentz contraction of the arms of the interferometer, also from that motion,
and (iii) the effects of the gas in the light paths which slightly slows the speed of light. In vacuum, that is with no gas present,
(i) and (ii) exactly cancel, but
 in the presence of a gas this cancellation effect is only partial and a small residual effect  occurs, which we now know explains
why the gas-mode interferometer experiments, from 1887 onwards, have always shown small rotation-induced fringe shifts. 

This
explanation was confirmed by analysing data from three other interferometer experiments, by Illingworth (1927), Joos (1930) and by
Jaseja {\it et al} (1964), that used He in the first two, and a He-Ne gas mixture in the last, allowing the effect of the
gas, in terms of its refractive index, to be demonstrated  by comparison with the air-mode data.  To show that this analysis of the
gas-mode interferometer was correct the results of the analysis were compared with the results from the 1st order in $v/c$ RF
travel-time coaxial cable experiments of Torr and Kolen (1981) and  DeWitte (1991).  

The key relevant aspect that arises from these
interferometer experiments is that of the Fitzgerald-Lorentz contraction of the arms.  Here that is a real physical effect, as
originally proposed by Fitzgerald and Lorentz in the 19$^{th}$ century. In contrast in the spacetime ontology interpretation by
Minkowski and Einstein this contraction is merely a perspective effect, depending on the `viewpoint' of an observer.  But the above
experimental data has being showing all along that the contraction was  physical  with its magnitude determined by the speed of motion
of the arms through a physically existing 3-space, where as usual the contraction is in the direction of motion only.  Such a uniform 
speed of itself has no connection with gravity. The observed speed is simply that of the apparatus through space, and in principle the
experimentalists could choose that speed.  So the contraction effect is caused by motion relative to a substratum, with apparently the
contraction arising from the interaction between the atoms forming the arms being affected by that uniform motion. 

So the argument is
that a 3-space exists, and  has structure, although we have as yet no measure of the size or nature of that structure, and that the
amalgamation of the geometric models of time and 3-space into a four dimensional spacetime was not mandated by experiment.  As well
the velocity field formalism in (\ref{eqn:CG3}) is Galilean covariant,  which means that observers in relative motion may transform
the velocity field using a Galilean transformation.  This is not in contradiction with the Lorentz transformation; these two
transformation rules relate the same  data but in different forms.  Hence the above suggests that the observed motion and the
contraction effect are the consequence of a substructure to space itself, and not some flowing particulate matter.   But then gravity
turns out to be merely a consequence of the space itself being non-static and non-uniform, that is when its structure  is in relative
motion, This means that the structure in one region of space is moving relative to the structure in a different region of space, so the
motion as such is only ever a differential motion, never a motion relative to some global background, whereas with a particulate
interpretation of the flow, the motion would have to be relative to some background geometry, and we would be back to the original
dualistic  aether theories.  

The relative motion of space itself is dramatically illustrated by the so-called Lense-Thirring effect. This is really the consequence
of vorticity in the flow, that is, one region of space is rotating relative to a neighbouring region of space \cite{GPB}. This is to be
detected by the gyroscopes aboard the Gravity Probe B satellite experiment.   There the spin direction of the gyroscopes is simply
carried  by the locally rotating space, with that rotation measured by comparison with distant space using light from 
a distant star. This vorticity or `frame-dragging' effect, as it is called in General Relativity, does not require any dynamical
calculation as would be the case if the vorticity was caused by some particulate matter  moving through
space. This vorticity is produced by the earth by means of its  rotation,  and as well  its  linear motion, upon the
local space.  The smaller component of the space-vorticity  effect caused by
the earth's   rotation   has been determined  from the laser-ranged satellites
LAGEOS(NASA) and LAGEOS 2(NASA-ASI) \cite{Ciufolini}, and the data is agreement with the vorticity interpretation to within  
$\pm10\%$.  However that  experiment cannot detect the larger  component of the  vorticity  induced by the 
linear motion  of the earth
  as that effect is not cumulative, while the rotation induced component is
cumulative.

 Miller didn't use the above  theory for the interferometer, but used the changes  in the observed velocity over a
year to calibrate the instrument; that is, he detected the motion of the earth about the sun in a purely laboratory experiment. Of
course in doing so he also detected the rotation of the earth about its own axis, but not relative to the sun, rather relative to the
fixed stars; that is he saw a sidereal and not a solar day effect. A re-analysis of that data
\cite{Cahill2,RGC} using the above interferometer theory has shown that the data reveals not only the orbital speed of the earth about
the sun but an in-flow component towards the sun, in agreement with (\ref{eqn:A2}).

So the evidence is that space has a differentially moving substructure, but that this motion has no absolute meaning, that is the
motion of space is just that, and not the movement of some constituents located in a space. So it is space itself that {\it flows}. 
A simple analogy to help visualise this is to think of space as an abstract network of connected patterns, where the connections have
an approximate embedding in a geometrical 3-space, but that embedding does not  imply that the 3-space is a separate entity;
rather it is an approximate coarse-grained description of the connectivity of the patterns.  Then as these patterns evolve in time, as
a real process, by older connections disappearing, and new connections forming, we can talk about the motion of one part of the pattern
system moving relative to other parts, so long as there is sufficient continuity, over time, of the pattern connectivity.  These
patterns in turn may be explained as internal informational relations, as discussed in \cite{PSS,CahillBook}.

\vskip12pt
\section{\bf  Conclusion}
\vskip6pt 

Historically the phenomenon of gravity was first explained by Newton in terms of  a gravitational acceleration field.
Later Einstein proposed a geometric theory which explained gravity in terms of  curvature of a four-dimensional manifold. 
However as shown herein, both these formalisms, in the cases where they have been explicitly tested, may be re-written in
terms of a velocity field formalism, with the acceleration field given in terms of the Euler `fluid' acceleration, though
with vorticity and relativistic corrections.  That by itself is  remarkable, and shows that the nature of gravity may have
been misunderstood all along. But even more significant is that a unique generalisation to that velocity field formalism
introduces a  dynamical effect that successfully explains a variety of known `gravitational anomalies', the most
dramatic being the so-called `dark matter' effect seen in spiral galaxies.   The strength of the new spatial
self-interaction dynamics is found from experimental data to be determined by $\alpha$, the fine structure constant, at
least to within experimental errors.

The new theory of gravity is able to explain various gravitational anomalies.  
The theory describes gravity as an inhomogeneous in-flow,  whether into matter or into a central `attractor' which is a
purely dynamical quantum-space effect, and essentially reveals space to be a quantum-foam process, with the strength of the
self-interactions in this process set by the fine structure constant, while
$G$ specifies the strength of the effect of matter in producing the spatial in-flow. 
As reported in \cite{Cahill2} there is 
experimental evidence that the in-flow velocity field
is now evident in older experimental data, although not recognised as such by the experimentalists involved.  Both the in-flow
past the earth towards the sun, and also past the earth into the local galactic cluster are evident.  As well the in-flow equations display
turbulence, and this also is evident in older experimental data. This of course amounts to the discovery of a new form of gravitational wave,
which is unlike that predicted by the Einstein theory.  Hence there is in fact a great deal of experimental and
observational  evidence that demonstrates the success of the new theory of gravity. 

Given that there is then considerable evidence  that the velocity field formalism represents a significant development
in our understanding of gravity, the question then arises as to what interpretation we might consider.  This new theory of
gravity has been shown to involve the fine structure constant, but this does not mean that the flow equations are
themselves quantum-theoretic. Nevertheless that the fine structure constant arises in both the phenomenon of gravity and
also in atomic, molecular and elementary particle systems, suggests that we are seeing, for the first time, suggestions
of a grand unification of the, so far, disjointed phenomena that physicists have uncovered.  As discussed in \cite{PSS,CahillBook} a
new {\it information-theoretic} modelling of reality is under development, and there space and matter arise as
self-organising informational patterns,  where the `information' here refers to internal information, and not to observer
based information. There we see the first arguments that indicate the logical necessity for quantum behaviour, at both the
spatial level and at the matter level.  There space is, at one of the lowest levels, a quantum-foam system undergoing
ongoing classicalisation.  That model suggest that gravity is caused by matter changing the processing rate of the
informational system that manifests as space, and as a consequence space effectively  `flows' towards matter. However this
is not a `flow' of some form of `matter' through space, as previously considered in the aether models or in the
`random' particulate  Le Sage kinetic theory of gravity, rather the flow is an ongoing rearrangement of the quantum-foam
patterns that form space, and indeed only have a geometrical description at a coarse-grained level.   Then the `flow' in
one region is  relative only to the patterns in nearby regions, and not relative to some {\it a priori} background
geometrical space.   The classical description of that flow necessarily involves the Euler `fluid' acceleration, as only
that construction has the required covariance property, but then that requirement immediately requires Newton's inverse
square law in the special case of small test objects external to a large central spherically symmetric mass, as was the
case for the solar system.  So not only does the new theory of gravity explain  numerous anomalies, it also explains the
origin of Newton's famous law for gravity. But also, significantly, it shows that this law, even in the non-relativistic
limit,  is not always valid.  The assumption that the inverse square law was `universally' valid in the non-relativistic regime, of
course, led to the fruitless search for `dark matter'.  Even more significant is that the dark matter effect is not within General
Relativity; this is most easily seen by noting that the GR formalism contains only one parameter, namely $G$, and certainly not the
fine structure constant. This happened because GR was constructed to agree with Newtonian gravity in the non-relativistic limit, and
that theory is now seen to be deficient even in that limit.   

Theories must be tested by experiment, and a whole new field of experimentation is now
possible in which laboratory Cavendish  experiments  can be used to extract the value of $\alpha$, and as discussed herein
there is ample evidence that this is possible, and indeed is the explanation for the long-standing problem in accurately
measuring $G$. The new theory is then suggesting that these laboratory experiments are essentially
quantum gravity experiments, and that they are revealing highly significant signatures of a deep unification of physics,
namely the unification of gravitational theory with the quantum theory, and to do that we have to abandon not only
Newtonian gravity, but also General Relativity and its curved spacetime formalism, the latter being a highly mathematical
disguise for the  classical description of an underlying processing quantum-foam system.  This implies that quantum
gravity effects do not set in at the extremely small scales of the Planck length and time, but  manifest  already in
numerous laboratory experiments. As well, because $\alpha$ now occurs  in both  atomic  and gravitational physics it is
presumably necessary to consider that
$\alpha$ is a fundamental dimensionless quantity, characterising in both cases a common deep random process, for that is
the role that
$\alpha$ plays in QED, and that there the electronic charge is given by $e=\sqrt{\alpha\hbar c}$.

Further results from the new theory of gravity are in \cite{CahillDM}.

\vskip12pt
\section{\bf  Acknowledgements}
\vskip6pt 

Thanks to   Peter Morris and David Roscoe for  valuable comments and suggestions.

\end{document}